\documentclass[10pt,aps,prd,superscriptaddress,twocolumn,floatfix,tightenlines,showkeys,showpacs,nofootinbib]{revtex4-1}

\usepackage{amsmath}
\usepackage{amsfonts}
\usepackage{amsthm}
\usepackage{amssymb}
\usepackage{bbm}
\usepackage{slashed}
\usepackage{cancel}
\usepackage{dcolumn}
\usepackage{bm}
\usepackage{graphicx}
\usepackage{color}
\usepackage{url}
\usepackage[utf8]{inputenc}
\newcommand{\be}{\begin{equation}}
\newcommand{\ee}{\end{equation}}
\newcommand{\bea}{\begin{eqnarray}}
\newcommand{\eea}{\end{eqnarray}}
\newcommand{\ba}{\begin{eqnarray*}}
\newcommand{\ea}{\end{eqnarray*}}
\newcommand{\barr}{\begin{array}}       
\newcommand{\earr}{\end{array}}
\newcommand{\bwt}{\begin{widetext}}
\newcommand{\ewt}{\end{widetext}}
\mathchardef \mhyphen="2D
\def\nn{\nonumber}

\newcommand{\Eq}[1]{Eq.\,(\ref{#1})}
\newcommand{\Sec}[1]{Sec.\,\ref{#1}}
\newcommand{\Fig}[1]{Fig.\,\ref{#1}}

\newcommand{\Cit}[1]{\,\cite{#1}}

\DeclareMathOperator{\sgn}{sgn}
\newcommand{\Z}{\mathbb{Z}}

\def\nn{\nonumber}

\begin{document}

\title{Finite temperature Schwinger pair production in coexistent electric and magnetic fields }

\author{Mrunal Korwar}
\email{mrunal.korwar@students.iiserpune.ac.in}
\affiliation{Indian Institute of Science Education and Research, Homi Bhabha road, Pashan, Pune 411008, India.}
\affiliation{Department of Physics, University of Wisconsin-Madison, Madison, WI 53706, USA.}
\author{Arun M. Thalapillil}
\email{thalapillil@iiserpune.ac.in}
\affiliation{Indian Institute of Science Education and Research, Homi Bhabha road, Pashan, Pune 411008, India.}

\date{\today}

\begin{abstract}
We compute Schwinger pair production rates at finite temperature, in the presence of homogeneous, concurrent electric and magnetic fields. Expressions are obtained using the semiclassical worldline instanton formalism, to leading order, for spin-$0$ and spin-$\frac{1}{2}$ particles. The derived results are valid for weak coupling and fields. We thereby extend previous seminal results in the literature, to coexistent electric and magnetic fields, and fermions. 
\end{abstract}

\maketitle

\section{Introduction}
The non-perturbative pair production of electrically and magnetically charged particles in the background of large field strengths has garnered much interest and study over the years. Sauter\Cit{Sauter:1931zz}, as well as Heisenberg and Euler\Cit{Heisenberg1936}, had speculated that sufficiently large electric fields could lead to spontaneous pair production of $e^+$- $e^-$. The notion was further sharpened and investigated comprehensively by Schwinger\Cit{Schwinger:1951nm}; deriving the imaginary part of the QED one-loop effective action. These results were then further generalised by various authors to diverse cases -- for instance, to extended objects such as magnetic monopoles\Cit{Affleck:1981ag}, spatial or temporal inhomogeneous fields\Cit{DunneWorldline, DunneFluctuationPre} and arbitrary gauge couplings\Cit{AFFLECK,Gould}, to cite a few examples (see for instance\Cit{Dunne:2004nc, Ruffini:2009hg} and references therein for a more complete discussion). Exact analytic expressions are known nevertheless only for a few special cases and extending investigations into hitherto unexplored regimes is an ongoing endeavour.

The worldline path integral formalism has proven to be a potent method for perturbative and non-perturbative quantum field theoretic computations. The origins of the method may be traced to ideas by Fock\Cit{Fock:1937dy}, Nambu\Cit{Nambu:1950rs} and the Feynman worldline representation of one-loop effective actions\Cit{Feynman:1950ir,Feynman:1951gn}. The formalism was, for instance, leveraged to compute pair-production rates for magnetic monopoles at strong coupling\Cit{Affleck:1981ag,AFFLECK}. With the development of string theoretic techniques towards understanding gauge theory scattering amplitudes\Cit{Halpern:1977he,Halpern:1977ru,Polyakov:1987ez,Bern:1991aq,Strassler:1992zr}, the method found further resurgence and applications (see for example\Cit{Schubert:2001he} and related references); particularly, in our context, conveniently accommodating computations with large external fields\Cit{Schmidt:1993rk,Shaisultanov:1995tm,Adler:1996cja, Reuter:1996zm,Bastianelli:2004zp}. 

Among the pertinent extensions to non-perturbative pair production rates at zero temperature, are the inclusion of finite temperature corrections. This has received much attention in the literature\Cit{Dittrich, Cox:1984vf, Selivanov:1986tu, LoeweRojas,Elmfors:1993wj, Ganguly,Hallin,ELMFORS1995141,Gies:1998vt,Gies:1999vb,Gavrilov:2007hq, Gavrilov:2007ij,Kim:2010qq,King:2012kd, Brown:2015kgj, Medina,Gould,Gould:2018ovk}. There has been some discussion and disagreement in the literature though, over these thermal corrections, particularly in the constant electric field case lately\Cit{Brown:2015kgj,Medina,Gould,Gould:2018ovk}. Thermal corrections for this case was recently computed\Cit{Medina} and extended to arbitrary coupling\Cit{Gould,Gould:2018ovk}, using worldline path integral techniques. 

Our aim in this work is to extend these results to the case when there are homogeneous (spatially and temporally) electric and magnetic fields simultaneously present. We compute leading order thermal corrections, using worldline path integral techniques, to the non-perturbative vacuum decay rates when there are coexistent electric and magnetic fields. We work in a regime where the coupling constant is small, and the external fields are also relatively weak. As far as we know, these expressions have not been computed before in the literature. We will largely follow techniques developed in\Cit{Affleck:1981ag,DunneWorldline, DunneFluctuationPre,Schubert:2001he, Medina}. In the limit of vanishing temperature ($T\rightarrow 0$), one recovers the well-known results in literature\Cit{Schwinger:1951nm, Nikishov:1969tt, 1970SPhD...14..678B,Popov:1971iga, Daugherty:1976mg, Cho:2000ei, KimDONPageEparallelB}. When the magnetic field vanishes ($B\rightarrow 0$), in the case of scalar quantum electrodynamics (SQED), the results are seen to relapse into the known expressions for pure homogeneous electric fields, computed recently\Cit{Medina}. In quantum electrodynamics (QED), with fermions, we also obtain new expressions in the $B\rightarrow 0$ limit that complement these recent SQED results. 

It is well known that even at zero temperature ($T=0$), the presence of a magnetic field parallel to the electric field ($E \shortparallel B$), leads to interesting modifications to vacuum decay rates, relative to the pure electric field case. The vacuum decay rates, per unit volume, at $T=0$ for homogenous $E \shortparallel B$ are given by\Cit{Schwinger:1951nm, Nikishov:1969tt, 1970SPhD...14..678B,Popov:1971iga, Daugherty:1976mg, Cho:2000ei, KimDONPageEparallelB}
 \bea
 \label{eq:ebsppt0}
\Gamma_{T=0, \text{\tiny{scalar}}}^{E \shortparallel B} &=& \sum_{k=1}^{\infty}\frac{(-1)^{k+1}q^{2}EB}{8\pi^{2}k\sinh(k\pi B/E)} \exp\Big{[}-\frac{m^{2}k\pi}{q E}\Big{]} \; , ~~~~\\
\Gamma_{T=0, \text{\tiny{fermion}}}^{ E \shortparallel B} &=& \sum_{k=1}^{\infty}\frac{q^{2}EB\coth(k\pi B/E)}{4\pi^{2}k} \exp\Big{[}-\frac{m^{2}k\pi}{q E}\Big{]} \nn \; .
\eea
Here, $m$ and $q$ are the mass and electric charge of the particle under consideration. Note that in addition to the usual enhancement due to extra degrees of freedom in the spin-$\frac{1}{2}$ case, the vacuum decay rates in the fermion case may be further enhanced, relative to the scalar case, when $B > E$. 

Note also that, for any homogeneous $\vec{E}^{'}$ and $\vec{B}^{'}$ fields, for which the Lorentz invariant $\vec{E}^{'}\cdot \vec{B}^{'} \neq 0$, one may go to a frame of reference with boost ($\vec{\upsilon}$) given by\Cit{Landau:1982dva}
\be
\frac{\vec{\upsilon}}{1+\lvert\vec{\upsilon}\rvert^2}=\frac{\vec{E}^{'}\times \vec{B}^{'}}{\lvert \vec{E}^{'}\rvert^2+\lvert \vec{B}^{'}\rvert^2}\; ,
\ee
where the transformed fields ($\vec{E}$ and $\vec{B}$) are parallel to each other. This is the so-called centre-of-field frame. Since the vacuum decay rate per unit volume is a Lorentz invariant, one may conveniently compute it in this centre-of-field frame. The formulas for homogeneous $E \shortparallel B$ are therefore potentially of wide applicability.  For homogeneous fields with $\vec{E}^{'}\cdot \vec{B}^{'} = 0$, but the fields not equal in magnitude, a reference frame may be found where the transformed field is purely electric or magnetic\,\cite{Landau:1982dva}. In this latter scenario, the relevant expressions are those of single field Schwinger pair production.

Apart from being of significant theoretical interest, scenarios with parallel electric and magnetic fields are also relevant in various astrophysical systems. For instance, it is believed that neutron stars such as pulsars have strong electrical fields parallel to the magnetic field in their polar vacuum gap regions\Cit{2004hpa..book.....L}. Neutron star surface temperatures are expected to reach $\sim 10^5\,\mathrm{K}$. Non perturbative production of exotic states such as millicharged particles, which may form a component of dark matter, may occur in these vacuum gap regions and provide hitherto unknown constraints on these states\Cit{Korwar:2017dio} (in the context of constraints from non-perturbative production, in pure $E$ or $B$ fields, also see \Cit{Gies:2006hv} for millicharged particle bounds from accelerator cavities, and \Cit{Hook:2017vyc, Gould:2017zwi} for bounds on magnetic monopoles). These settings also, therefore, make the results phenomenologically very relevant.

In \Sec{sec:sqed}, we discuss the derivation in the case of SQED. Towards the exposition of necessary techniques and to fix notations, we re-derive the known zero temperature result for the case of $E \shortparallel B$ using worldline instantons, before presenting the main result for finite temperature. Then, in \Sec{sec:qed}, we consider QED. Results are presented for spin-$\frac{1}{2}$ particles in the zero temperature and finite temperature cases. The finite temperature SQED and QED Schwinger pair production results, for $E \shortparallel B$, are new and readily generalise earlier seminal results in the literature\Cit{Brown:2015kgj, Medina,Gould}. Even for the zero temperature cases, to the best of our knowledge, this is the first time that an explicit and complete derivation is being presented for vacuum decay rates, when $E \shortparallel B$, using worldline instanton techniques. We summarise our main results, shortcomings of the derivations and future directions in \Sec{sec:conc}.

\section{Thermal Pair production for $E \shortparallel B$ in SQED}{\label{sec:sqed}

We would like to calculate decay rates for vacua, made metastable by the presence of large external fields.
Let us denote the probability for vacuum to vacuum transitions by $\big{|}\big{<}0_{out}|0_{in}\big{>}\big{|}^{2}$. In presence of external fields sourced by a potential $A$, the probability for vacuum to vacuum transitions are given by
\be
\Big{<}0_{out}|0_{in}\Big{>}_{A} = \exp(iW^{\mathbb{M}}[A]) \; ,
\ee
where $W^{\mathbb{M}}[A]$ is the Minkowskian effective action for the theory under consideration. 

Expressed in terms of Euclidean quantities, specialising now to Scalar electrodynamics (SQED), we have explicitly
\be
\exp(-W^{\mathbb{E}}[A]) = \int \mathcal{D}\phi\, \mathcal{D}\phi^{*}~ \exp[-S_{\mathbb{E}}] \; .
\ee
Here,
\be
S_{\mathbb{E}} =  \int d^{4}x \, (\phi^{*}(-D^{2}+m^{2})\phi) + \frac{1}{4} F_{\mu\nu}^{2} \; .
\ee
In the above Euclidean expressions, the covariant derivative $D_{\mu} := (\partial_{\mu}+i q A_{\mu})$, external gauge field ${A}_{\mu}:=(A_{1},A_{2},A_{3},A_{4})$ with $A_{4}:=-iA_{0}$, and field tensor $F_{\mu\nu}:=\partial_{\mu}{A}_{\nu} - \partial_{\nu}{A}_{\mu}$. In terms of Euclidean quantities, the Lorentz invariant vacuum decay rate per unit volume is given by
\be
\Gamma_{\text{\tiny{VD}}} = 2\,\text{Im}\Big{(}\frac{W^{\mathbb{E}}[A]}{V_{4}^{\mathbb{E}}}\Big{)} \; .
\label{eq:vacdecayrate}
\ee

We simplify the effective action further, following a standard technique\Cit{AFFLECK,DunneWorldline}, and after performing a functional integration, one obtains
\be
2\text{Im}(W^{\mathbb{E}}[A]/V_{4}^{\mathbb{E}}) = 2\,\text{Im} (\text{Tr} \,\text{ln} (-D^{2}+m^{2})/V_{4}^{\mathbb{E}})\; .
\ee
Using Frullani's integral identity $\text{Tr}(\ln \mathcal{M})= -\int_{0}^{\infty} \frac{dz}{z} \, \text{Tr}(\exp{[-\mathcal{M} z]}-\exp{[-z]})$\Cit{Schwinger:1951nm, gradshteyn2007}, dropping terms that do not contribute to the imaginary part, and converting the trace to a path integral, leads then to the well-known expression for the SQED one-loop Euclidean effective action\Cit{AFFLECK,DunneWorldline}
\bea
W^{\mathbb{E}}[A] &=& -\int_{0}^{\infty}\frac{dz}{z} \exp{[-m^{2} z]} \oint_{x(0)=x(z)} \mathcal{D} x \nn \\
&& \exp\Big{[} - \int_{0}^{z} d\tau \big{[}\frac{1}{4} x^\prime{}^{2} + i q A^{\mu} x^\prime_{\mu}\big{]}\Big{]}  \; .
\eea
Here, $x^\prime$ denotes differentiation with respect to $\tau$. An implicit assumption that the coupling constant is small ($q^2 \ll 1$) has been made while writing the above result, by dropping non-local interaction terms that are higher order in the coupling constant. Now, making a substitution $\tau = z\,u$ and $z \rightarrow z/m^2 $ gives
\bea
W^{\mathbb{E}}[A] &=& -\int_{0}^{\infty}\frac{dz}{z} \exp{[-z]} \oint_{x(0)=x(1)} \mathcal{D} x \nn \\
&& \exp\Big{[}-\Big{(} \frac{m^2}{4z}\int_{0}^{1} du\dot{x}^2 + i q \int_{0}^{1}du A^\mu\dot{x}_\mu  \Big{)}\Big{]} \; .
\label{eq:effactzuvar}
\eea
$\dot{x}$ denotes differentiation with respect to $u$. 

Evaluating the $z$ integral above, gives
\bea
W^{\mathbb{E}}[A] &=& -2 \oint_{x(0)=x(1)} \mathcal{D} x~ \mathcal{K}_{0}\Big{(}m\big{(}\int_{0}^{1} \dot{x}^{2} du\big{)}^{1/2}\Big{)} \nn \\
&& \exp\Big{[}-i q \int_{0}^{1}du A^\mu \dot{x}_\mu \Big{]} \; ,
\eea
where $\mathcal{K}_{0}(z)$ is the modified Bessel function of the second kind. For $x\gg 1$, we have the asymptotic formula $ \mathcal{K}_{0}(x) \sim \exp(-x)$ and hence, when $m\sqrt{\int_{0}^{1} \dot{x}^{2} du} \gg 1$, the above expression may be simplified to
\bea
W^{\mathbb{E}}[A]&\simeq&-\sqrt{\frac{2\pi}{m}}\oint_{x(0)=x(1)} \mathcal{D} x \,\frac{1}{[\int_{0}^{1}\dot{x}^{2}\, du]^{1/4}} \nn \\
&& \exp\Big{[} -m\sqrt{\int_{0}^{1} \dot{x}^{2} \, du} - i q \int_{0}^{1} A^\mu \dot{x}_\mu \,du \Big{]} \; .~~~~~
\label{eq:effactapprox}
\eea
The assumption $m\sqrt{\int_{0}^{1} \dot{x}^{2}} \gg 1$ is equivalent to making a weak field approximation $q E/m^2 \ll 2 \pi$ \Cit{AFFLECK,DunneWorldline}. We will therefore also assume that the external electromagnetic fields are relatively weak and satisfy $q |\bar{\bar{F}}|/m^2 \lesssim 1$, for field strengths $|\bar{\bar{F}}|$. Finally, note that \Eq{eq:effactapprox} may equivalently be obtained by making a  saddle point approximation, to the $z$ integral in \Eq{eq:effactzuvar}.

Considering the terms in the exponent as part of an effective action,
\be
S_{\text{\tiny{eff}}} := m\sqrt{\int_{0}^{1} \dot{x}^{2} \, du} + i q \int_{0}^{1} A^\mu \dot{x}_\mu \,du \; ,
\label{eq:seff}
\ee
the corresponding Euler-Lagrange equations are
\be
m \ddot{x}_{\xi} = i q \sqrt{\int_{0}^{1}du \, \dot{x}^{2}}\,\,\, F_{\xi\zeta}\, \dot{x}^{\zeta} \; .
\ee
The antisymmetry of $F_{\mu\nu}$ immediately implies that
\be
\dot{x}^{2}=\rho^{2} \; ,
\ee
where $\rho$ is a constant.

Specialising now to temporally and spatially homogeneous $E\shortparallel B$, let us choose $\vec{E}$ and $\vec{B}$ in the $x_{3}$ direction, without loss of generality. We then have for the Field tensor $\bar{\bar{F}}$,
 \be
 F_{12}=-F_{21}=B;\, F_{34}=-F_{43}=iE \; .
\ee
 This leads to the equations of motion
 \bea
m\ddot{x}_{1} &=& iq\rho B\dot{x}_{2} ~, ~~ m\ddot{x}_{2} = -iq \rho B\dot{x}_{1} \;,\nn \\
m\ddot{x}_{3} &=& -q \rho E\dot{x}_{4}~,~~ m\ddot{x}_{4} = q \rho E\dot{x}_{3} \; .
\label{eq:sct0eqs}
\eea

To clarify ideas and general techniques, that shall be adopted in the finite temperature derivation, we first derive the well-known result in the $T=0$ case, using the worldline instanton formalism. Though this result is well-known and has been derived using many other techniques\Cit{Nikishov:1969tt, 1970SPhD...14..678B,Popov:1971iga, Daugherty:1976mg, Cho:2000ei, KimDONPageEparallelB}, we believe that a systematic derivation of this has not been presented before in the literature, using worldline path integral methods.

We note from \Eq{eq:sct0eqs} that the equations of motion for $x_{1},x_{2}$ and $x_{3},x_{4}$ are decoupled from each other. The set of equations for $x_{1},x_{2}$ give rise to hyperbolic solutions, which fail to satisfy the periodic boundary condition $x_{\mu}(0)=x_{\mu}(1)$, as required by \Eq{eq:effactzuvar}. Thus, the only solutions for $x_{1}$ and $x_{2}$ are trivial solutions. For $x_3$ and $x_4$ one finds solutions 
\be
x_{3}=\frac{m}{qE}\cos\Big{(}\frac{qE \rho u}{m}\Big{)} ~,~~ x_{4}=\frac{m}{qE}\sin\Big{(}\frac{qE \rho u}{m}\Big{)}\; ,
\label{eq:eomsolt0}
\ee
satisfying the required periodic boundary conditions. Let us collectively denote these solutions by $\bar{x}$. Note that in the above, one must have $\rho=2\pi k R=\frac{m2 k \pi}{qE}$, to satisfy the boundary conditions. These solutions therefore represent a circle in the $x_{3}-x_{4}$ plane, with radius $R= m/qE$. This is equivalent to the situation in the pure $E$ case\Cit{AFFLECK}. The effective action, with these solutions ($\bar{x}$), is then given by
\be
S_{\text{\tiny{eff}}}(\bar{x})=\frac{m^2 k \pi}{qE} \; .
\ee

Let us now compute the fluctuation prefactor for this solution(for general techniques, see for instance \Cit{Weinberg:2012pjx, Marino:2015yie, Paranjape:2017fsy}). To leading order, the fluctuation prefactor is proportional to $\sim \text{det}[\delta^2 S_{\text{\tiny{eff}}}/\delta x_{\nu} \delta x_{\mu}]^{-1/2}$, evaluated at the solutions to the equations of motion, with appropriate boundary conditions.

Define the prefactor matrix at zero temperature
\bea
\mathcal{P}^{0,\text{\tiny{scalar}}}_{\mu\nu} &:=& \frac{\delta^2 S_{\text{\tiny{eff}}}}{\delta x_{\nu}(u') \delta x_{\mu}(u)}\Bigg{|}_{\bar{x}}   \\ 
&=& -\Bigg{[}\frac{qE \delta_{\mu\nu}}{2k\pi}\frac{d^{2}}{du^{2}} - iqF_{\mu\nu}\frac{d}{du}\Bigg{]}\delta(u-u') \nn \\
&-& \frac{2k\pi q E}{R^{2}} \bar{x}_{\mu}(u) \bar{x}_{\nu}(u') \; . \nn
\eea
The relevant determinant, with zero modes removed, may be expressed using the matrix determinant lemma (see for instance\Cit{Harville, Medina}) as
\bea
\label{eq:prefactsqed}
\text{det}'[\mathcal{P}^{0,\text{\tiny{scalar}}}] &=& \text{det}'[\mathcal{C}_0] \Big{[}1-2k\pi  \frac{q E}{R^2} \\ 
&& \int{\int{du\,du' \bar{x}_{\mu}(u)\,(\mathcal{C}^{'\,-1}_0)_{\mu\nu}\,\bar{x}_{\nu}(u')}}\Big{]}(-2 k \pi q E) \nn
\eea
$\mathcal{C}^{'\,-1}_0:=\mathcal{G}^0(u,u')$ is to be interpreted as a Green's function. In the $E \shortparallel B$ case, we have
\be
\mathcal{C}'_0 :=
\begin{bmatrix}
-\frac{q E}{2k\pi}\frac{d^{2}}{du^{2}} & iq B\frac{d}{du} & 0 & 0 \\
-iq B\frac{d}{du}      & -\frac{q E}{2k\pi}\frac{d^{2}}{du^{2}} & 0 & 0 \\
0 & 0 & -\frac{q E}{2k\pi}\frac{d^{2}}{du^{2}} & - q E\frac{d}{du} \\
0 & 0 &  q E\frac{d}{du} & -\frac{ q E}{2k\pi}\frac{d^{2}}{du^{2}}\\
\end{bmatrix} \; .
\ee
Two of the eigenvalues are --- $2\pi q E(l^{2}/k-l)$, corresponding to eigenvectors $(0,0,\cos(2l\pi u),\sin(2l\pi u))$ and $(0,0,\sin(2l\pi u),-\cos(2l\pi u))$, and $2\pi q E(l^{2}/k+l)$, corresponding to eigenvectors $(0,0,\sin(2l\pi u),\cos(2l\pi u))$ and $(0,0,\cos(2l\pi u),-\sin(2l\pi u))$. The other two eigenvalues have the form --- $2\pi q E\big{(}-\frac{i B l}{E} + \frac{l^{2}}{k} \big{)}$, corresponding to eigenvectors $(1, i ,0 ,0)\exp[2\pi i l u]$ and $(i,-1,0,0)\exp[2\pi i l u]$, and $2\pi q E\big{(}\frac{i B l}{E} + \frac{l^{2}}{k} \big{)}$, corresponding to eigenvectors $(1, -i ,0 ,0)\exp[2\pi i l u]$ and $(i,1,0,0)\exp[2\pi i l u]$. In all cases $l$ runs from $1$ to $\infty$. With these, one obtains
\bea
 \text{det}'[\mathcal{C}_0] &=& \Big{(}\frac{(4\pi N_{0})^{2}}{m^{4}}\Big{)}^{2} \frac{1}{(2k\pi q E)^{2}}\prod_{l\neq 0 ,k}\frac{\big{(}(l^{2}/k-l)\big{)}^{2}}{l^{4}/k^{2}} \nn \\
 && \prod_{l\neq 0 , l=-\infty}^{\infty} \frac{\Big{[}2\pi q E(\frac{l^{2}}{k} - \frac{iB l}{E})\Big{]}^{2} }{(2\pi q E)^{2} l^{4}/k^{2}} \;,
\eea
where $N_{0} = m^{2}k\pi/q E$. The infinite products may be simplified\Cit{gradshteyn2007}, and one obtains the compact expression

\be
\text{det}'[\mathcal{C}_0]=\Big{(}\frac{8k\pi^{3} (-1)^{k+1}}{q^3 E^3}\Big{)}^{2} \Big{(}\frac{E\sinh(k \pi B/E)}{k \pi B}\Big{)}^{2} \; .
\ee
It is interesting to compare this to the equivalent expression in the case of pure $E$\Cit{DunneFluctuationPre}.

The only part remaining to be calculated is the non-local factor that appears in \Eq{eq:prefactsqed} ---  $[1-2 k\pi  \frac{q E}{R^2}\int{\int{du\,du' \bar{x}_{\mu}(u)\,(\mathcal{G}^0)_{\mu\nu}\,\bar{x}_{\nu}(u')}}]$. Here, for the non-trivial solutions, the only relevant part of $\mathcal{C}'_0$ is the $(3-4)$ block. The Green's function $\mathcal{G}^0(u,u')$ can be obtained in the standard way by constructing a spectral representation. Utilising the relation $\mathcal{C}'{}^{-1}_0 = V \mathcal{C}_{0, D}^{'\,-1} V^{-1}$,  with $V$ the eigenvector column matrix and $\mathcal{C}'_{0,D}$ the diagonal matrix,	 one gets
\bea
\mathcal{G}^0(u,u') &=& \sum_{\substack{l\neq 0\\ l \neq k \\ l=-\infty}}^{\infty}\frac{1}{2\pi q E(l^{2}/k -l)} \\
&&\left(\begin{array}{cc}
\cos\left[2\pi l\left(u-u'\right)\right] & -\sin\left[2\pi l\left(u-u'\right)\right]\\
\sin\left[2\pi l\left(u-u'\right)\right] & \cos\left[2\pi l\left(u-u'\right)\right]
\end{array}\right) \nn \; .
\eea
With the non-trivial solutions for $x_{3}$ and $x_{4}$ this gives
\be
\int_{0}^{1}du\int_{0}^{1}du'\,\bar{x}_{\mu}\left(u\right)\left(\mathcal{G}^0\right)_{\mu\nu}\bar{x}_{\nu}\left(u'\right)=0 \;.
\ee
Therefore, due to the decoupling in \Eq{eq:sct0eqs} leading to trivial solutions for $x_1$ and $x_2$, the non-local part of the prefactor matrix determinant comes out to be unity, in complete analogy to the pure $E$ case\Cit{Medina}.

Putting all the factors together, the fluctuation prefactor for fixed $k$ finally comes out to be
\be
\mathcal{F}_{T=0, \text{\tiny{scalar}}}^{E \shortparallel B} = \frac{V_{4}^{\mathbb{E}}(-1)^{k+1}q^{2}E^{2}i}{16\pi^{3}k^{2}}\frac{k\pi B}{E\sinh(k\pi B/E)} \; .
\ee

The relevant part of the SQED Euclidean effective action then becomes
\be
 W_{T=0, \text{\tiny{scalar}}}^{\mathbb{E}, E \shortparallel B}= \sum_{k=1}^{\infty} \frac{iV_{4}(-1)^{k+1}q^{2}EB}{16\pi^{2}k\sinh(k\pi B/E)} \exp\Big{[}-\frac{m^{2}k\pi}{qE}\Big{]} \; .
\ee  

From this, using \Eq{eq:vacdecayrate}, the $T=0$ vacuum decay rate per unit volume, in SQED for homogeneous $E\shortparallel B$, may be calculated finally as
\be
\Gamma_{T=0, \text{\tiny{scalar}}}^{E \shortparallel B}= \sum_{k=1}^{\infty}\frac{(-1)^{k+1}q^{2}EB}{8\pi^{2}k\sinh(k\pi B/E)} \exp\Big{[}-\frac{m^{2}k\pi}{q E}\Big{]} \;
\label{eq:sqedt0vd}
\ee
This matches the well-known zero temperature SQED expression in literature\Cit{Schwinger:1951nm, Nikishov:1969tt, 1970SPhD...14..678B,Popov:1971iga, Daugherty:1976mg, Cho:2000ei, KimDONPageEparallelB}, as given in \Eq{eq:ebsppt0}. Note that it also reduces to the pure $E$ case in the limit $B\rightarrow 0$, as expected.

With this warm-up derivation in the zero temperature case, clarifying ideas and techniques, we now proceed to thermal Schwinger pair production in SQED when one has homogeneous $E\shortparallel B$ fields. For calculating finite temperature vacuum decay rates, for scalar particles in the presence of a homogeneous electromagnetic field, we need to calculate the imaginary part of the SQED thermal effective action. The supplemental requirement in the thermal case is that the Euclidean time direction must now be compact with endpoints identified and one  requires $x_{4}(1)\equiv x_{4}(0)+n\beta$\Cit{Weinberg:2012pjx, Marino:2015yie, Paranjape:2017fsy, McKeon:1992if, McKeon:1993sh, Shovkovy:1998xw}, with $n\in \mathbb{Z}$. Here, $\beta^{-1}$ is the temperature ($T$), that is assumed to be much less than the mass of the particle under consideration ($T \ll m$).

The SQED Euclidean effective action at finite temperature is given by
\bea
W^{\mathbb{E}}_{T\neq 0, \text{\tiny{scalar}}}&=& \sum_{n\in \Z}-\sqrt{\frac{2\pi}{m}}\oint_{\substack{x_{4}(1)\equiv x_{4}(0)+n\beta \\ x(0)=x(1)}} \mathcal{D} x \frac{1}{[\int_{0}^{1}\dot{x}^{2}\, du]^{1/4}} \nn \\
&& \exp\Big{[} -m\sqrt{\int_{0}^{1} \dot{x}^{2} du} - iq\int_{0}^{1} A^\mu \dot{x}_\mu du \Big{]}
\eea
One has again assumed weak fields, $q \bar{\bar{F}}/m^2 \ll 1$, and small couplings $q^2 \ll 1$. Note that $n=0$ coincides with the expression already derived, for zero temperature. We focus on the $n\neq 0$ contributions. The terms in the exponent above, are again to be considered as part of some effective action ($S_{\text{\tiny{eff}}}$).

To find the relevant thermal instantons, we need to find solutions to the equations of motion \Eq{eq:sct0eqs}, that are now additionally compact in $x_{4}$, with period $n\beta$\Cit{Weinberg:2012pjx, Marino:2015yie, Paranjape:2017fsy, McKeon:1992if, McKeon:1993sh, Shovkovy:1998xw}. Thus, we have to essentially find local sections of the zero temperature instanton solutions \Eq{eq:eomsolt0}, that are additionally periodic by $n\beta$ in the $x_4$ direction. For such viable solutions to exist, we must have $2R\geq n\beta$, as is clear from geometry. This implies a bound $n_{max}=\lfloor2R/\beta\rfloor$, where $\lfloor x\rfloor$ denotes the integer less than or equal to $x$. This means that there are no one-loop thermal contributions for $T< q E/2m\equiv T_{*}$, defining a critical temperature $T_{*}$ for a given mass $m$ and charge $q$. Since there are no thermal corrections below $T=T_{*}$, it may provide a partial resolution with some earlier studies\Cit{Cox:1984vf, ELMFORS1995141, Gies:1998vt}, where it was argued that there are no thermal corrections at one-loop (also see discussions in \Cit{Medina, Gould:2018ovk}).

Now, for $n\in \mathbb{Z}^{-}$, i.e. solutions satisfying the boundary condition
\be
x_{4}(1)=x_{4}(0)+n\beta ~;~~n\in \mathbb{Z}^{-} \; ,
\ee		
there are two solutions (see \Fig{fig:FiniteTemperatureInstanton-}). For the smaller path ($I^{-}$), subtending angle $\theta_{n}$ at the center, $\Theta' = 2\pi k+\theta_{n}$ is the total angle subtended by $k$ windings. The explicit solution ($\bar{x}^{\tiny{\text{T}}, I^{-}}$) in this case is given by
\be
x_{3} = R\cos(\Theta' u + \pi - \theta_{n}/2) ~,~~ x_{4}=R\sin(\Theta' u + \pi - \theta_{n}/2) \; ,
\ee		
with the end-points of $x_4$ identified. There are again no non-trivial solution for $x_{1}$ and $x_{2}$ satisfying the requisite periodic boundary conditions, similar to the zero temperature case. The corresponding effective action may be computed for this solution, from \Eq{eq:seff}, and comes out to be
\bea
S_{\text{\tiny{eff}}}(\bar{x}^{\tiny{\text{T}}, I^{-}})&=&mR\Theta' - mR\frac{\Theta'}{2} + \frac{m^{2}}{2eE} \sin(\theta_{n})  \\
&=& \frac{m^{2}}{2qE}\Big{[}2\pi k + 2\arcsin\big{(}\frac{nT_{*}}{T}\big{)}\Big{]} \nn \\
&&~~~~~~+ \frac{nm}{2T}\sqrt{1-\frac{n^{2}T_{*}^{2}}{T^{2}}} \nn \; ,
\eea
where $R=m/q E$ and $T_{*}=qE/2m$. The relation between angle subtended $\theta_{n}$ and temperature $T$ is
\be
\sin\Big{(}\frac{\theta_{n}}{2}\Big{)} = -\frac{n\beta}{2R} = -\frac{n T_{*}}{T}~~, \, n\in \mathbb{Z}^{-} \; .
\label{eq:angleTrel}
\ee
\begin{figure}
\centering
\includegraphics[width=9.0cm]{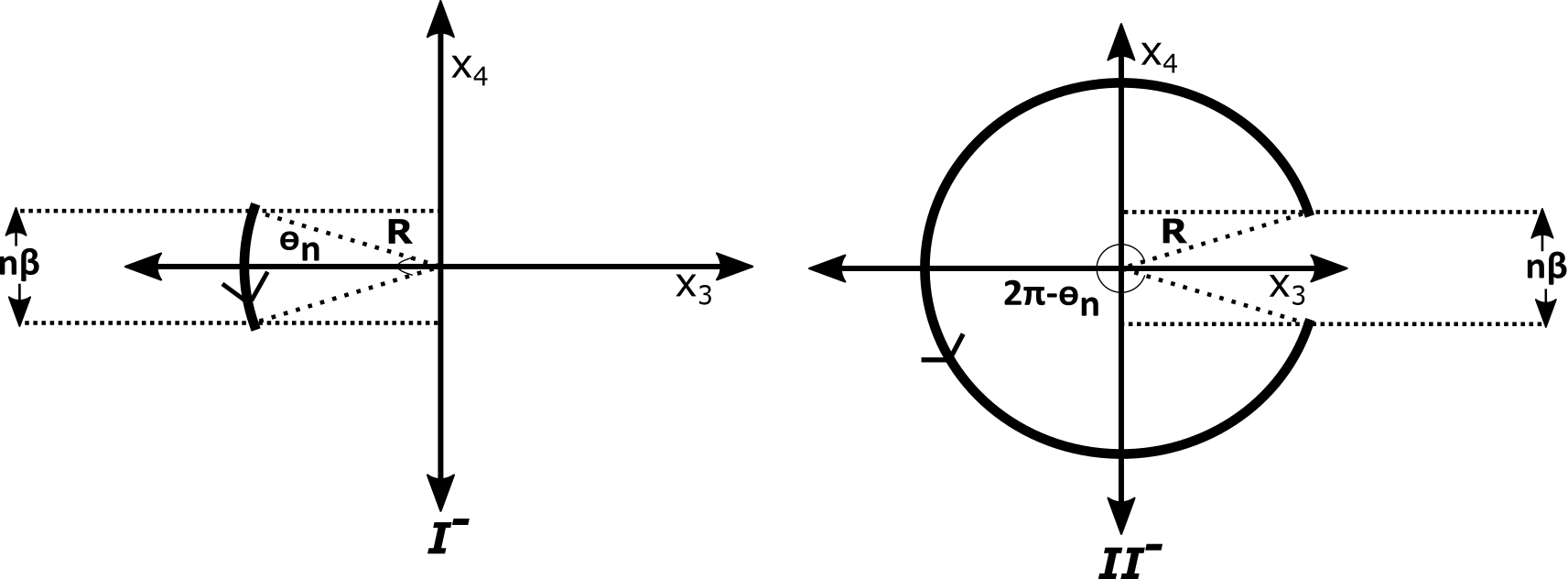}
\caption{The $I^-$ (left) and $II^-$ (right) solutions corresponding to $n\in \mathbb{Z}^{-}$. The short path $I^-$ does not contribute an imaginary part to the Euclidean effective action, while the long path $II^-$ does. It is therefore the $\bar{x}^{\tiny{\text{T}}, II^{-}}$ solution that would contribute to vacuum decay rates.}
\label{fig:FiniteTemperatureInstanton-} 
\end{figure}

As we shall see, a calculation of the fluctuation prefactor for the $I^{-}$ solution ($\bar{x}^{\tiny{\text{T}}, I^{-}}$) shows that it does not contribute to the imaginary part of the Euclidean effective action. Therefore, the solution $\bar{x}^{\tiny{\text{T}}, I^{-}}$ may only contribute to the free energy, and there is no contribution to the vacuum decay rate from it.

For the longer path ${II}^{-}$, shown in \Fig{fig:FiniteTemperatureInstanton-}, subtending angle $2\pi - \theta_{n}$ at the centre, $\Theta = 2\pi (k+1)-\theta_{n}$ is the total angle subtended by $k$ windings. The non-trivial part of the solution $\bar{x}^{\tiny{\text{T}}, II^{-}}$ is given by
\be
x_{3}(u) = R\cos(\Theta u + \theta_{n}/2) ~,~~ x_{4}(u) = R\sin(\Theta u + \theta_{n}/2) \; .
\label{eq:longminus}
\ee
The corresponding effective action, using \Eq{eq:seff}, may be calculated and gives
\bea
\label{eq:seffiim}
S_{\text{\tiny{eff}}}(\bar{x}^{\tiny{\text{T}}, II^{-}}) &=& mR\Theta - mR\frac{\Theta}{2} - \frac{m^{2}}{2qE} \sin(\theta_{n}) \\
&=& \frac{m^{2}}{2qE}\Big{[}2\pi (k+1) + 2\arcsin\big{(}\frac{nT_{*}}{T}\big{)}\Big{]} \nn \\
&&~~~~~~ + \frac{nm}{2T}\sqrt{1-\frac{n^{2}T_{*}^{2}}{T^{2}}} \nn \; ,
\eea
where, as before, $T_{*}=qE/2m$ and $R= m/qE$. The relation between $\theta_{n}$ and $n$ for ${II}^{-}$, is same as in \Eq{eq:angleTrel}. This solution, as we shall demonstrate while calculating the fluctuation prefactor, will be one that does contribute to the vacuum decay rate, by giving an imaginary part to the Euclidean effective action.

For the positive integer case, $n\in \mathbb{Z}^{+}$ case, we have the requirement 
\be
x_{4}(1)=x_{4}(0)+n\beta  ~;~~ n\in \mathbb{Z}^{+} \; .
\ee
There are again two solutions (see \Fig{fig:FiniteTemperatureInstanton+}). For the smaller path ($I^{+}$), subtending angle $\theta_{n}$ at the center, $\Theta' = 2\pi k+\theta_{n}$ is the total angle subtended. $k$ is again the number of windings. As is amply clear from \Fig{fig:FiniteTemperatureInstanton+} and geometry, the explicit solution ($\bar{x}^{\tiny{\text{T}}, I^{+}}$) for this case is
\be
x_{3} = R\cos(\Theta' u - \theta_{n}/2) ~,~~ x_{4}=R\sin(\Theta' u - \theta_{n}/2) \; .
\ee
These solutions give	for the effective action
\bea
S_{\text{\tiny{eff}}}(\bar{x}^{\tiny{\text{T}}, I^{+}}) &=& mR\Theta' - mR\frac{\Theta'}{2} + \frac{m^{2}}{2qE} \sin(\theta_{n})  \\
&=& \frac{m^{2}}{2 q E}\Big{[}2\pi k + 2\arcsin\big{(}\frac{nT_{*}}{T}\big{)}\Big{]} \nn \\
&&~~~~~~+\frac{nm}{2T}\sqrt{1-\frac{n^{2}T_{*}^{2}}{T^{2}}} \nn \; ,
\eea
and as in the $n\in \mathbb{Z}^{-}$ case the computation of prefactor shows that it only contributes to the free energy and not to pair production.

Coming now to the longer path ($II^{+}$), subtending an angle $2\pi - \theta_{n}$ at the center (see \Fig{fig:FiniteTemperatureInstanton+}) we have for k-windings, a total angle subtended $\Theta = 2\pi (k+1)-\theta_{n}$. The $II^{+}$ solution, similar to $II^{-}$ of \Eq{eq:longminus}, will contribute to the imaginary part of the effective action. This solution ($\bar{x}^{\tiny{\text{T}}, II^{+}}$) is explicitly 
\be
x_{3}(u) = R\cos(\Theta u +\pi +\theta_{n}/2) ~,~~ x_{4}(u) = R\sin(\Theta u +\pi+ \theta_{n}/2) \; .
\ee
The corresponding effective action is calculated to be
\bea		
\label{eq:seffiip}
S_{\text{\tiny{eff}}}(\bar{x}^{\tiny{\text{T}}, II^{+}}) &=& mR\Theta - mR\frac{\Theta}{2} - \frac{m^{2}}{2eE} \sin(\theta_{n})  \\
&=& \frac{m^{2}}{2qE}\Big{[}2\pi (k+1) - 2\arcsin\big{(}\frac{nT_{*}}{T}\big{)}\Big{]} \nn \\
&&~~~~~~ -\frac{nm}{2T}\sqrt{1-\frac{n^{2}T_{*}^{2}}{T^{2}}} \nn \; .
\eea
$T_{*}$, $R$, are as defined earlier and the relation between $\theta_{n}$ and $n$ is now
\be
\sin\Big{(}\frac{\theta_{n}}{2}\Big{)} = \frac{n\beta}{2R} = \frac{n T_{*}}{T}~~, \, n\in \mathbb{Z}^{+} \; .
\label{eq:angleTrelpos}
\ee

Note from \Eq{eq:seffiim} and \Eq{eq:seffiip} that the two solutions, $\bar{x}^{\tiny{\text{T}}, II^{-}}$ and $\bar{x}^{\tiny{\text{T}}, II^{+}}$ contributing to the vacuum decay rate, actually give equivalent expressions for the exponential factor. The contribution to pre-exponential factors will also be seen to be similar, for both solutions. Hence, the full sum over $n\in \mathbb{Z}$ may be replaced just by twice sum over $n\in \mathbb{Z}^{+}$. Hence, from now on, we will just consider the solution $\bar{x}^{\tiny{\text{T}}, II^{+}}$ for presenting the relevant calculations. 
\begin{figure}
\centering
\includegraphics[width=9.0cm]{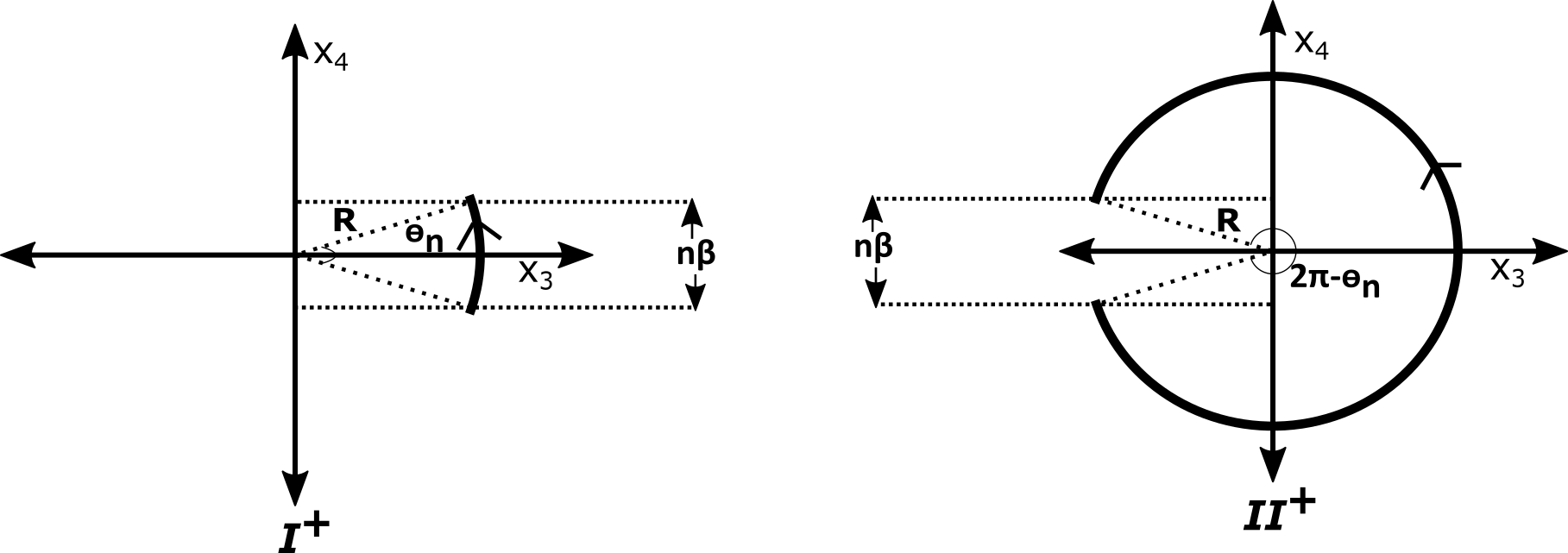}
\caption{The $I^+$ and $II^+$ solutions corresponding to $n\in \mathbb{Z}^{+}$. The short path $I^+$, as in the earlier case, does not contribute an imaginary part to the Euclidean effective action. The long path $II^+$ does contribute an additional negative eignevalue from the non-local part in the prefactor matrix, and hence contributes to an imaginary part for the Euclidean effective action. It is thus the $\bar{x}^{\tiny{\text{T}}, II^{+}}$ solution again that would contribute to vacuum decay rates we are interested in.}
\label{fig:FiniteTemperatureInstanton+}
\end{figure}
		
Let us now compute the fluctuation prefactor relevant to the $\bar{x}^{\tiny{\text{T}}, II^{+}}$ solution. Again, define a prefactor matrix
\bea
\mathcal{P}^{T,\text{\tiny{scalar}}}_{\mu\nu}&:=&\frac{\delta^2 S_{\text{\tiny{eff}}}}{\delta x_{\nu}(u') \delta x_{\mu}(u)}\Bigg{|}_{\bar{x}^{\tiny{\text{T}}, II^{+}}}   \\
&=&  -\Big{[}\frac{qE\delta_{\mu\nu}}{\Theta}\frac{d^2}{du^2} - iqF_{\mu\nu}\frac{d}{du}\Big{]}\delta(u-u') \nn \\
&&-\frac{\Theta qE{\bar{x}^{\tiny{\text{T}}, II^{+}}_{\mu}}(u){\bar{x}^{\tiny{\text{T}}, II^{+}}_{\nu}}(u')}{R^{2}} \; .\nn
\eea
The relevant determinant, with the zero modes removed, may be written as\Cit{Harville}
\bea
\label{eq:sqedtdetp}
\text{det}'[\mathcal{P}^{T,\text{\tiny{scalar}}}]&=&\text{det}'[\mathcal{C}_T]\Big{[}1-\Theta \frac{q E}{R^2} \\
&& \int{\int{du\,du' \bar{x}^{\tiny{\text{T}}, II^{+}}_{\mu}(u)\,(\mathcal{C}^{'\,-1}_T)_{\mu\nu}\,\bar{x}^{\tiny{\text{T}}, II^{+}}_{\nu}(u')}}\Big{]} \nn\;.
\eea
Here, $\mathcal{C}^{'\,-1}_T:=\mathcal{G}^T(u,u')$ is again to be interpreted as an appropriate Green's function, without zero modes. The matrix $\mathcal{C}'_T $ is given in this case by
\be
\mathcal{C}'_T :=\begin{bmatrix}
	 -\frac{qE}{\Theta}\frac{d^{2}}{du^{2}} & iqB\frac{d}{du} & 0 & 0 \\
	 -iqB\frac{d}{du}       & -\frac{qE}{\Theta}\frac{d^{2}}{du^{2}} & 0 & 0 \\
	 0 & 0 & -\frac{q E}{\Theta}\frac{d^{2}}{du^{2}} & -q E\frac{d}{du} \\
	 0 & 0 &  q E\frac{d}{du} & -\frac{q E}{\Theta}\frac{d^{2}}{du^{2}}
	 \end{bmatrix}\; .
\ee
Note the presence of additional elements in the $1 \mhyphen 2$ block, depending on magnetic field strength $B$, compared to the equivalent matrix in the pure electric field case\Cit{Medina}.

For calculating $\text{det}'[\mathcal{C}_T]$, we utilise the result\Cit{Marino:2015yie, Gelfand:1959nq, Levit:1976fv, Levit:1977ed}
\be
 \Bigg{|} \frac{\text{det}'[\mathcal{C}_T]}{\text{det}'[\overline{\mathcal{C}}_T]}\Bigg{|} = \Bigg{|}\frac{\prod_\alpha \xi_{\alpha}}{\prod_\alpha{\bar{\xi}}_{\alpha}}\Bigg{|}=\Bigg{|}\frac{\text{det}\, \zeta_{\nu}^{\, (a)}(1)}{\text{det}\, \bar{\zeta}_{\nu}^{\,(a)}(1)}\Bigg{|} \; ,
 \label{eq:gymethod}
\ee
where $\overline{\mathcal{C}}_T$ is the matrix formed from $\mathcal{C}_T$ by excluding all non-diagonal terms. $\xi_{\alpha}$ and $\bar{\xi}_{\alpha}$ are the eigenvalues of $\mathcal{C}_T$ and $\overline{\mathcal{C}}_T$ respectively. The matrices $\zeta_{\nu}^{\,(a)}(u)$ and $\bar{\zeta}_{\nu}^{\,(a)}(u)$ satisfy the following set of equations\Cit{Marino:2015yie, Gelfand:1959nq, Levit:1976fv, Levit:1977ed}
\bea
\mathcal{C}_T{}_{\mu\nu}~\zeta_{\nu}^{\,(a)}(u) &=& 0 \,;\,\, \zeta_{\nu}^{\,(a)}(0)=0 \,;\,\, \dot{\zeta}_{\nu}^{\,(a)}(0)=\delta_{\nu}^{\,a} \nn \\
\overline{\mathcal{C}}_T{}_{\mu\nu}~\bar{\zeta}_{\nu}^{\,(a)}(u) &=& 0 \,;\,\, \bar{\zeta}_{\nu}^{\,(a)}(0)=0 \,;\,\, \dot{\bar{\zeta}}_{\nu}^{\,(a)}(0)=\delta_{\nu}^{\,a} \; . 
\eea
\\
Since the eigen spectrum for $\mathcal{C}_T$ is unknown, we may use the second equality in terms of the $\zeta$ and $\bar{\zeta}$ matrices to calculate $\text{det}'[\mathcal{C}_T]$\Cit{Marino:2015yie, Gelfand:1959nq, Levit:1976fv, Levit:1977ed}. For the homegenous $E\shortparallel B$ case we are considering, the coresponding $\zeta$ matrix, with appropriate boundary conditions, comes out to be
\bwt
\bea
\zeta (u) = 
\begin{bmatrix}
\frac{E}{\Theta B}\sinh\big{(}\frac{B\Theta u}{E}\big{)} & \frac{i E}{\Theta B}\Big{(}\cosh\big{(}\frac{B\Theta u}{E}\big{)}-1\Big{)} & 0                       & 0  \\
\frac{i E}{\Theta B}\Big{(}1-\cosh\big{(}\frac{B\Theta u}{E}\big{)}\Big{)} & \frac{E}{\Theta B}\sinh\big{(}\frac{B\Theta u}{E}\big{)} & 0                       & 0   \\
0 & 0 & \frac{\sin{\Theta u}}{\Theta}      & \frac{(-1+\cos{\Theta u})}{\Theta} \\
0 & 0 & \frac{(1-\cos{\Theta u})}{\Theta} & \frac{\sin{\Theta u}}{\Theta}
\end{bmatrix} \; .
\eea
\ewt

From this, we have at $u=1$ the determinant, $\text{det}[{\zeta}(1)] = \frac{4E^{2}}{\Theta^{4}B^{2}}(\cosh(B\Theta/E)-1)(1-\cos{\Theta})$. This is always positive. The $\bar{\zeta}(u)$ matrix comes out to be --- $u \cdot\mathbbm{1}_{4\times 4}$. Hence, the relevant determinant is just $\text{det}[\bar{\zeta}(1)] = 1$. Putting all the above results together and using \Eq{eq:gymethod}, $\sqrt{\text{det}'[\mathcal{C}_T]}$ may now be readily computed in our case as
\bea
\sqrt{\text{det}'[\mathcal{C}_T]} &=& N_T (-1)^{k} \Big{(}\frac{2\pi \Theta}{q E} \Big{)}^{2} \sqrt{\frac{2(1-\cos\Theta)}{\Theta^{2}}}  \\
&& \sqrt{\frac{2E^{2}(\cosh[B\Theta/E]-1)}{\Theta^{2}B^{2}}} \nn \; .
\eea
$N_T$ is a normalization factor that may be fixed explicitly by considering $\overline{\mathcal{C}}_T$ and the free theory. The factor $(-1)^{k}$ is related to the Morse index\Cit{morse1934calculus, Levit:1976fv,Paranjape:2017fsy} of the corresponding solution. 

The non-local part of the prefactor matrix determinant in \Eq{eq:sqedtdetp}, is of the form
\bea
\text{det}'[\mathcal{P}^{T,\text{\tiny{scalar}}}]&\supset& \Big{[}1-\Theta \frac{q E}{R^2}\int \int du\,du' \bar{x}^{\tiny{\text{T}}, II^{+}}_{\mu}(u)\,\nn \\
&&~~\mathcal{G}^T_{\mu\nu}\,\bar{x}^{\tiny{\text{T}}, II^{+}}_{\nu}(u') \Big{]}
\eea
$\mathcal{G}^T_{\mu\nu}$ are Green's functions satisfying  
\be
(\mathcal{C}'_T)_{\mu\beta}~\mathcal{G}^T_{\beta\nu}(u,u') = \delta_{\mu\nu}\,\delta(u-u') \; ,
\label{eq:gtfunc}
\ee
with vanishing Dirichlet boundary conditions. Since $x_{1}$ and $x_{2}$ do not have non-trivial solutions, satisfying required boundary conditions, the combinations containing $\mathcal{G}^T_{11}, \mathcal{G}^T_{12}, \mathcal{G}^T_{21}$, and $\mathcal{G}^T_{22}$ in the integral all trivially give zero. The remaining terms are those with $\mathcal{G}^T_{33}, \mathcal{G}^T_{34}, \mathcal{G}^T_{43}$ and $\mathcal{G}^T_{44}$. These are related to the $3\mhyphen 3$, $3\mhyphen 4$ and $4\mhyphen 4$ elements of $\mathcal{C}_T$, which only depend on the electric field $E$. This immediately suggests that the Green's function should match that computed in the pure $E$, thermal case\Cit{Medina}. Since $(\mathcal{C}_T)_{33}=(\mathcal{C}_T)_{44}$ and $(\mathcal{C}_T)_{43}=-(\mathcal{C}_T)_{34}$, it may be shown that $\mathcal{G}^T_{33}=\mathcal{G}^T_{44}$ and $\mathcal{G}^T_{43}=-\mathcal{G}^T_{34}$. Solving \Eq{eq:gtfunc}, considering cases $u>u'$ and $u<u'$, give 
\bwt
\bea
\mathcal{G}^T_{33} &=&\frac{1}{qE}\Big{[}  \sin\left( \frac{\Theta (u+u')}{2}\right) \cos\left( \frac{\Theta (u-u')}{2}\right) -\frac{\sin\left(\Theta|u-u'|\right)}{2}-\frac{2\sin\left(\frac{\Theta}{2}u\right)\sin\left(\frac{\Theta}{2}u'\right)\cos\left(\frac{\Theta}{2}\left(u-u'\right)\right)}{\tan\left(\frac{\Theta}{2}\right)}\Big{]} \\
\mathcal{G}^T_{43} &=& \frac{1}{qE}\Big{[}  \sin\left( \frac{\Theta (u+u')}{2}\right)\sin\left(\frac{\Theta(u-u')}{2}\right)+\frac{\sgn\left[u-u'\right]}{2}\big{(}\cos\left(\Theta|u-u'|\right)-1\big{)}-\frac{\sin\left(\Theta\left(u-u'\right)\right)+\sin\left(\Theta u'\right)-\sin\left(\Theta u\right)}{2 \tan\left(\frac{\Theta}{2}\right)}\Big{]} \; . \nn
\eea
\ewt
These are in agreement with the expressions found in\Cit{Medina}, for the pure E case. Putting all the above results together, the contribution of the non-local part, for $T\neq 0$ and homegeneous $E\shortparallel B$, come out to be
\be
\text{det}'[\mathcal{P}^{T,\text{\tiny{scalar}}}] \supset \frac{\Theta}{2}\cot\Big{(}\frac{\Theta}{2}\Big{)} \; .
\ee

This is manifestly negative, giving an extra negative mode for longer paths ($II^{\pm}$), and thus contributing an imaginary part to the Euclidean effective action $W_{T\neq0, \text{\tiny{scalar}}}^{\mathbb{E}, E \shortparallel B}$. Note that this is because the fluctuation prefactor is proportional to $\sim \text{det}[\delta^2 S_{\text{\tiny{eff}}}/\delta x_{\nu} \delta x_{\mu}]^{-1/2}$, evaluated at the stationary solutions. As alluded to before, the longer path solutions, therefore, contribute to vacuum decay rates. In contrast, substituting $\Theta'$ corresponding to the shorter paths ($I^{\pm}$), in place of $\Theta$, would give a non-local contribution which is positive. This finally makes the fluctuation prefactor real and hence contributes only to free energy. In the pure electric field case, this was checked by matching the $E\rightarrow 0$ limit of the short-path expressions\Cit{Medina}, with the exact free energy density of a non-interacting relativistic particle\Cit{Meisinger:2001fi}, when $\beta \rightarrow \infty$. A derivation of the free energy density using the standard proper time representation of the effective potential\Cit{Gies:1998vt}, in an external electric field, also matches that derived from the short-path expression. There is nevertheless some disagreement in the literature regarding the appropriate choice of path\Cit{Brown:2015kgj,Medina,Gould,Gould:2018ovk}. 

Taking all the contributions into account, the thermal SQED fluctuation prefactor, for fixed $k$, comes out to be
\bea	 
\mathcal{F}_{T, \text{\tiny{scalar}}}^{E \shortparallel B} &=& (-1)^{k}\frac{iV_{3}\beta}{4}\frac{q^{2}EB}{(2\pi)^{3/2}(nm\beta)^{1/2}\Theta\sinh\big{(}\frac{\Theta B}{2E}\big{)}} \nn \\
&& \Big{[}1-\Big{(}\frac{n\beta q E}{2m}\Big{)}^{2}\Big{]}^{-1/4} 
\eea

Finally, combining all the exponential and pre-exponential factors, the thermal vacuum decay rate per unit unit volume, to leading order, in the background of homogeneous $E \shortparallel B$ comes out to be
\be
\Gamma_{T\neq 0, \text{\tiny{scalar}}}^{E \shortparallel B} = \Gamma_{T= 0, \text{\tiny{scalar}}}^{E \shortparallel B} + \Gamma_{T, \text{\tiny{scalar}}}^{E \shortparallel B} H(T-T_{*})\; ,
\ee
with $\Gamma_{T= 0, \text{\tiny{scalar}}}^{E \shortparallel B}$ given by \Eq{eq:sqedt0vd} and $\Gamma_{T, \text{\tiny{scalar}}}^{E \shortparallel B}$ by
\bwt
\bea
\Gamma_{T, \text{\tiny{scalar}}}^{E \shortparallel B} = \sum_{n=1}^{n_{max}} \sum_{k=0}^{\infty}&\frac{ (-1)^{k} q^{2}EB}{(2\pi)^{\frac{3}{2}}(nm\beta)^{\frac{1}{2}}\Theta\sinh(\frac{\Theta B}{2E})}  \big{[}1-\big{(}\frac{n\beta q E}{2m}\big{)}^{2}\big{]}^{-\frac{1}{4}} \exp\Big{[}-\frac{m^{2}}{2 q E}\big{[}2\pi (k+1) - 2\arcsin\big{(}\frac{nT_{*}}{T}\big{)}\big{]} +\frac{nm}{2T}\sqrt{1-\frac{n^{2}T_{*}^{2}}{T^{2}}}\Big{]} \; .~~~~~~
\eea
\ewt
In above, $n_{max}=\lfloor2R/\beta\rfloor$, $H(x)$ is the Heaviside step function, and $\Theta = 2\pi (k+1)-\theta_{n}=2\pi(k+1)-2\arcsin(\frac{nT_{*}}{T})$. In the limit of $B\rightarrow 0$, $\Gamma_{T\neq 0, \text{\tiny{scalar}}}^{E \shortparallel B}$ reduces to the known expression for $\Gamma_{T\neq 0, \text{\tiny{scalar}}}^{E}$\Cit{Medina}. Also, note that when $T<T_{*}\equiv q E/2m $, the periodic boundary conditions on $x_4$ cannot be satisfied and there are no thermal corrections. In this case the result relapses to the zero temperature expression.

\section{Thermal Pair production for $E \shortparallel B$ in QED}{\label{sec:qed}

We now proceed to compute the non-perturbative pair-production rates for spin-$\frac{1}{2}$ particles in Quantum Electrodynamics (QED). The derivation is analogous to the SQED derivation, with some subtleties coming from the additional Pauli spin term and the necessities of fermionic functional integrations.

For QED, the Euclidean effective action for fermion field $\Psi$ is given by
\be
\exp(-W^{\mathbb{E}})= \int \mathcal{D}\Psi \mathcal{D}\bar{\Psi} \, \exp[-S_{\mathbb{E}}] \; ,
\ee
with
\be
S_{\mathbb{E}} = \int d^{4}x \,\bar{\Psi}(\slashed{D}+m)\Psi + \frac{1}{4} F_{\mu\nu}^{2} \; .
\ee
Here, we define $\slashed{D}=\gamma^{\mu}_{\mathbb{E}}D_{\mu}=\gamma^{\mu}_{\mathbb{E}}(\partial_{\mu}+iqA_{\mu})$ and $\bar{\Psi} = \Psi^{\dagger}\gamma^{4}_{\mathbb{E}}$. We have defined ${A}_{\mu}=(A_{1},A_{2},A_{3},A_{4})$ as before such that $A_{4}=-iA_{0}$ and $F_{\mu\nu}=\partial_{\mu}{A}_{\nu} - \partial_{\nu}{A}_{\mu}$. $\gamma^{\mu}_{\mathbb{E}}$ are Euclidean gamma matrices, which are related to the Minkowskian  gamma matrices through the relations
\be
 \gamma^{4}_{\mathbb{E}}= \gamma^{0}_{\mathbb{M}} ~,~~ \gamma^{i}_{\mathbb{E}} = -i\gamma^{i}_{\mathbb{M}} \; ,
 \ee
 in our convention. They satisfy
\be
 \{\gamma^{\mu}_{\mathbb{E}},\gamma^{\nu}_{\mathbb{E}}\}  = 2\delta^{\mu\nu} \, ,\, \gamma^{5}_{\mathbb{E}} = -\gamma^{1}_{\mathbb{E}}\gamma^{2}_{\mathbb{E}}\gamma^{3}_{\mathbb{E}}\gamma^{4}_{\mathbb{E}} \, ,\, \{\gamma^{5}_{\mathbb{E}},\gamma^{\mu}_{\mathbb{E}}\} = 0 \; .
\ee
For brevity, henceforth we will remove the subscript ($\mathbb{E}$) from the Euclidean gamma matrices. 

Performing the fermion functional integral gives
\be
W^{\mathbb{E}} = -\frac{1}{2} \text{Tr}\, \text{ln} [-D^{2} + m^{2} +\frac{1}{2}q\,\sigma_{\xi\zeta}F^{\xi\zeta}] \; ,
\ee
where $\sigma_{\mu\nu}=-\frac{i}{2}[\gamma_{\mu},\gamma_{\nu}]$. Using the Frullani integral identity\Cit{Schwinger:1951nm, gradshteyn2007}, this may be expressed  as
\be
W^{\mathbb{E}} =  \frac{1}{2} \int_{0}^{\infty}\frac{dz}{z}\text{Tr}\Big{\{}\exp{\Big{[}-z\Big{(}-D^{2}+m^{2} +\frac{1}{2}q\,\sigma_{\mu\nu}F^{\mu\nu}\Big{)}\Big{]}}\Big{\}} \; .
\ee
Note the additional factor of $1/2$ compared to the scalar case as well as the additional Pauli spin term. These lead to interesting differences between the SQED and QED results. Introducing fermionic coherent states\Cit{DHoker:1995uyv, DHoker:1995aat, Schubert:2001he} and simplifying, the above Euclidean effective action may be re-written as 
\bwt
\be
W^{\mathbb{E}}=\frac{1}{2}\int_{0}^{\infty}\frac{dz}{z} \exp(-m^{2}z) \oint_{x(0)=x(z)} \mathcal{D} x \exp\Big{[}-\int_{0}^{z}d\tau\Big{(}\frac{x^\prime{}^2}{4}+i q A^{\mu}x^\prime_{\mu}\Big{)}\Big{]}\text{Tr}_{f}\Big{\{}\exp\Big{[}-\frac{1}{2}z\, q \sigma^{\mu\nu}F_{\mu\nu}\Big{]}\Big{\}} \; .
\label{eq:qedeeffact}
\ee
\ewt
 In above, $\text{Tr}_{f}$ denotes a fermionic trace and we have assumed $q^2 \ll 1$. In terms of fermionic coherent states ($\eta$), this one-loop QED Euclidean effective action explicitly takes the form\Cit{Schubert:2001he}
 \bea
 W^{\mathbb{E}} &=& \frac{1}{2}\int_{0}^{\infty}\frac{dz}{z} e^{-m^{2}z}
\oint_{x(0)=x(z)} \mathcal{D} x \exp\Big{[}-\int_{0}^{z} d\tau\big{(}\frac{x^\prime{}^2}{4} + \nn \\ 
&& i q x^\prime_{\mu}A^{\mu}\big{)}\Big{]} \oint_{\eta(0)=-\eta(z)}\mathcal{D} \eta \exp\Big{[}-\int_{0}^{z}d\tau\big{(}\frac{\eta^{\mu} \eta^\prime_{\mu}}{2} - \nn \\
&& iq \eta^{\mu}F_{\mu\nu}\eta^{\nu}\big{)}\Big{]} \; .
 \eea
 Let us define
 \bea
 J &:=& \frac{1}{2}\int_{0}^{\infty}\frac{dz}{z} \, e^{-m^{2}z} \,
\oint_{x(0)=x(z)} \mathcal{D} x \nn \\
&& \exp\Big{[}-\int_{0}^{z} d\tau\big{(}\frac{x^\prime{}^2}{4}+i q x^\prime_{\mu}A^{\mu}\big{)}\Big{]} \; ,
 \eea
 and also note that 
 \be
 \oint_{\eta(0)=-\eta(z)} \mathcal{D} \eta \exp\left[-\int_{0}^{z}d\tau\, \eta \eta^\prime/2\right] =2^{d_{\mathbb{E}}/2}\; .
 \ee
 $d_{\mathbb{E}}$ is the number of Euclidean dimensions. Following a standard technique\Cit{Corradini:2015tik}, let us then re-write the Euclidean effective action as
\be
W^{\mathbb{E}} = 4~J~ \text{det}^{1/2}\Big{(}\mathbbm{1}-2i q \bar{\bar{F}}\big{(}\frac{d}{d\tau}\big{)}^{-1}\Big{)} \; .
\ee
As before, $\bar{\bar{F}}$ is the electromagnetic field tensor with components $F_{12}=-F_{21}=B$ and $F_{34}=-F_{43}=iE$. Note that $\text{det}^{1/2}(\mathbbm{1}-2i q \bar{\bar{F}}(\frac{d}{d\tau})^{-1})$ should be a Lorentz scalar. Hence it should depend on $\bar{\bar{F}}^{2}$ and hence only on the coupling constant as $q^{2}$\Cit{Corradini:2015tik}. Thus, we may relate $\text{det}^{1/2}(\mathbbm{1}-2i q \bar{\bar{F}}(\frac{d}{d\tau})^{-1}) =\text{det}^{1/2}(\mathbbm{1}+2i q \bar{\bar{F}}(\frac{d}{d\tau})^{-1})$. From this, we can write
\bea
Z^{2} &:=& \text{det}\Big{(}\mathbbm{1}-2i q \bar{\bar{F}}\big{(}\frac{d}{d\tau}\big{)}^{-1}\Big{)} \cdot 
\text{det}\Big{(}\mathbbm{1}+2i q \bar{\bar{F}}\big{(}\frac{d}{d\tau}\big{)}^{-1}\Big{)} \; , \nn \\
&=&  \text{det}\Big{(}\mathbbm{1}+4 q^{2}\bar{\bar{F}}^{2}\big{(}\frac{d}{d\tau}\big{)}^{-2}\Big{)} \; .
\eea
Using these definitions,
\begin{equation}
W^{\mathbb{E}} =4~J~Z^{1/2} \; .
\end{equation}

In the case of interest, we have
\be
\bar{\bar{F}}^{2}= \text{diag}(-B^{2}, -B^{2}, E^{2}, E^{2}) \; .
\ee
Since $\bar{\bar{F}}^{2}$ is diagonal, the factor $Z$ may be evaluated readily as
\bea
Z^{2}&=&\text{det}\Big{(}\text{diag}\big{[}1-4B^{2}q^{2}(d/d\tau)^{-2},\, 1-4B^{2}q^{2}(d/d\tau)^{-2}, \nn \\
&&\, 1+4E^{2}q^{2}(d/d\tau)^{-2}, \, 1+4E^{2}q^{2}(d/d\tau)^{-2}\big{]}\Big{)} \; .
\eea
From this, we find
\begin{equation}
Z= \text{det}\big{(}1-4B^{2}q^{2}(d/d\tau)^{-2}\big{)}\, \text{det}\big{(}1+4E^{2}q^{2}(d/d\tau)^{-2}\big{)} \; .
\label{eq:zvals}
\end{equation}
The above determinant may be obtained in the usual way, by solving the eigenvalue problem 
\be
-\frac{d^{2}}{ds^{2}} f(s)= \lambda f(s) \; ,
\ee
with anti-periodic boundary condition $f(z)=-f(0)$. The eigenfunctions satisfying these boundary conditions are 
\bea
f_{(1)}(s) &=& \cos(2\pi(t+1/2)s/z) \; ,\nn \\
 f_{(2)}(s)&=& \sin(2\pi(t+1/2)s/z) ~;~~ t=0,1,\cdots \infty \;.
\eea
The corresponding eigenvalues are given by
\be
\lambda_{t}= \frac{(2\pi(t+1/2))^{2}}{z^{2}} \; .
\ee

Substituting this in \Eq{eq:zvals}, and taking into account the two-fold degeneracy, we get after a simplification of the infinite products\Cit{gradshteyn2007},
\bea
Z&=&\Big{[}\prod_{t=0}^{\infty}\Big{(}1+\frac{4B^{2}q^{2}}{\lambda_{t}}\Big{)}\Big{]}^{2}\Big{[}\prod_{t'=0}^{\infty}\Big{(}1-\frac{4E^{2}q^{2}}{\lambda_{t'}}\Big{)}\Big{]}^{2}  \; ,\nn \\
&=& \cosh^{2}(q Bz)\cos^{2}(q Ez) \; .
\eea
Substituting these results back, one obtains
\bea
W^{\mathbb{E}}&=& 2\int_{0}^{\infty}\frac{dz}{z} \, e^{-m^{2}z} \,
\oint_{x(0)=x(z)} \mathcal{D} x  \exp\Big{[}-\int_{0}^{z} d\tau\big{(}\frac{x^\prime{}^2}{4}\nn \\
&&~~~~~~+ i q x^\prime_{\mu}A_{\mu}\big{)}\Big{]} \cosh( q Bz)\cos( q Ez) \; .
\label{eq:qedeeffactsimpl}
\eea

We now make a change of variable $\tau\rightarrow zu$, $z\rightarrow z/m^{2}$, as before, and perform the $z$ integral using a saddle point approximation. The additional cosine term, in the fermion case above, gives an imaginary part in the exponential and hence does not modify the saddle point\Cit{DunneWorldline}. Also, the hyperbolic cosine term when written in its exponential form contributes a factor $\pm q Bz/m^{2}$ to the integrand's exponent. In the limit of weak fields, $q |\bar{\bar{F}}|/m^2 \ll 1$, this does not modify the saddle point either. Hence the saddle point for the $z$ integral turns out to be $z_{0} = \frac{m}{2}(\int_{0}^{1} \dot{x}^{2}\, du)^{1/2}$. 

The relevant one-loop Euclidean effective action in QED is then given by
\bea
W^{\mathbb{E}} &=&  2\sqrt{\frac{2\pi}{m}}\oint_{x(0)=x(1)} \mathcal{D} x \frac{1}{[\int_{0}^{1}\dot{x}^{2}\, du]^{1/4}} \nn \\
&& \exp\Big{[} -m\sqrt{\int_{0}^{1} \dot{x}^{2} du}\,\, - \,\,i q \int_{0}^{1} A.\dot{x} du \Big{]} \nn \\
&& \cos \Big{[}\frac{q E z_{0}}{m^{2}}\Big{]}\cosh\Big{[}\frac{q B z_{0}}{m^{2}}\Big{]} \; .
\eea

Considering the terms in the exponent above as part of an effective action, the corresponding Euler-Lagrange equations are again given by
\begin{equation}
 m \ddot{x}_{\xi} = i q \sqrt{\int_{0}^{1}du \, \dot{x}^{2}}\,\,\, F_{\xi\zeta} \dot{x}^{\zeta}  \; .
\end{equation}

Let us initially consider the $T=0$ case, as before. For $E\shortparallel B$, $E$ and $B$ assumed to be in the $x_{3}$ direction, there are again no non-trivial solutions for $x_{1}$ and $x_{2}$, satisfying the periodic boundary conditions. Hence, in complete analogy to SQED, the only non-trivial solutions are
\be
x_{3}=R \cos(2 k \pi u) ~,~~x_{4}=R \sin(2 k \pi u) \; .
\ee
This leads to the effective action and the exponential part of the QED vacuum decay rate
\be
S_{\text{\tiny{eff}}}(\bar{x}(u))=\frac{m^2 k \pi}{q E} \; .
\ee

The additional factors, for $z_{0} = m \rho/2 = m^{2}k\pi/q E$, come out to be
\be
\cos(q E z_{0}/m^{2})\cosh(q B z_{0}/m^{2})\rightarrow (-1)^{k}\cosh(k\pi B/E) \; .
\ee
The fluctuation prefactor,for fixed $k$, is then
\bea
\mathcal{F}_{T=0, \text{\tiny{fermion}}}^{E \shortparallel B} &=& -2\cdot\mathcal{F}_{T=0, \text{\tiny{scalar}}}^{E \shortparallel B}  \\
&=& -2\cdot \frac{V_{4}^{\mathbb{E}}(-1)^{k+1}q^{2}E^{2}i}{16\pi^{3}k^{2}}\frac{k\pi B}{E\sinh(k\pi B/E)} \; . \nn
\eea
Combining everything, the one-loop Euclidean effective action is
\be
W_{T=0, \text{\tiny{fermion}}}^{\mathbb{E}, E \shortparallel B}= \sum_{k=1}^{\infty} \frac{iV_{4}^{\mathbb{E}}q^{2}EB}{8\pi^{2}k} \exp\Bigg{[}-\frac{m^{2}k\pi}{qE}\Bigg{]}\coth[k\pi B/E] \; ,
\ee
giving the vacuum decay rate in QED at zero temperature,
\be
\Gamma_{T=0, \text{\tiny{fermion}}}^{ E \shortparallel B}= \sum_{k=1}^{\infty}\frac{q^{2}EB\coth(k\pi B/E)}{4\pi^{2}k} \exp\Big{[}-\frac{m^{2}k\pi}{q E}\Big{]} \; .
\label{eq:qedsppt0}
\ee
This expression, derived using the worldline path integral method, matches the familiar zero temperature QED expression\Cit{Schwinger:1951nm, Nikishov:1969tt, 1970SPhD...14..678B,Popov:1971iga, Daugherty:1976mg, Cho:2000ei, KimDONPageEparallelB}
, as given in \Eq{eq:ebsppt0}.

Let us now turn to the finite temperature case ($T\neq 0$). The computation follows the zero temperature case largely, with few additional complexities introduced by the requirement of the periodicity criteria along $x_4$\Cit{Weinberg:2012pjx, Marino:2015yie, Paranjape:2017fsy, McKeon:1992if, McKeon:1993sh, Shovkovy:1998xw}, as in SQED. To compute the fermion pair production at finite temperature, we again must consider solutions that are compact in the $x_4$ direction, with end-points identified, and separated by $n\beta$. Based on \Eq{eq:qedeeffact} and \Eq{eq:qedeeffactsimpl}, the one-loop effective action for fermions at finite temperature is
\bwt
\bea
W_{T\neq0, \text{\tiny{fermion}}}^{\mathbb{E}, E \shortparallel B} &=&\sum_{n\in \Z}2\sqrt{\frac{2\pi}{m}}\oint_{\substack{x_{4}(1) = x_{4}(0)+n\beta \\ x(0) = x(1)}} \mathcal{D} x \frac{1}{[\int_{0}^{1}\dot{x}^{2}\, du]^{1/4}} \exp\Bigg{[} -m\sqrt{\int_{0}^{1} \dot{x}^{2} du} - iq\int_{0}^{1} A.\dot{x} du \Bigg{]} \cos \Big{[}\frac{ q E z_{0}}{m^{2}}\Big{]}\cosh\Big{[}\frac{qB z_{0}}{m^{2}}\Big{]}\;.~~~
\eea
\ewt

For $E\shortparallel B$, in the $q \bar{\bar{F}}/m^2 \ll 1$ regime, the equations of motion do not change compared to the corresponding scalar case. Hence, nor does the value of $S_{\text{\tiny{eff}}}(\bar{x}^{\tiny{\text{T}}, II^{+}})$, computed earlier in \Eq{eq:seffiip}. This leads to an exponent with $S_{\text{\tiny{eff}}}(\bar{x}^{\tiny{\text{T}}, II^{+}})$, which using $z_{0} = mR\Theta/2$ for $T\neq 0$ leads to a factor
\bea
&&\exp(-S_{\text{\tiny{eff}}}(\bar{x}^{\tiny{\text{T}}, II^{+}})) \cos\Big{[}\frac{q E z_{0}}{m^{2}}\Big{]}\cosh\Big{[}\frac{q B z_{0}}{m^{2}}\Big{]}\rightarrow \nn \\
&&\exp\Big{[}-\frac{m^{2}}{2 q E}\Big{[}2\pi (k+1) - 2\arcsin\big{(}\frac{nT_{*}}{T}\big{)}\Big{]} +\frac{nm}{2T}\sqrt{1-\frac{n^{2}T_{*}^{2}}{T^{2}}}\Big{]}\nn \\
&& \cos\Big{(}\frac{\Theta}{2}\Big{)}\cosh\Big{(}\frac{B\Theta}{2E}\Big{)}
\eea

The determinant of the prefactor matrix $\text{det}'[\mathcal{P}^{T,\text{\tiny{fermion}}}]$, which appears in the computation of the QED fluctuation prefactor, also mostly remains the same as in the thermal SQED case. The relevant fluctuation prefactor hence becomes
\bea	
\mathcal{F}_{T\neq 0, \text{\tiny{fermion}}}^{E \shortparallel B}&=& -2\,\mathcal{F}_{T \neq 0, \text{\tiny{scalar}}}^{E \shortparallel B} \\
&=& -2\cdot (-1)^{k}\frac{iV_{3}\beta}{4}\frac{q^{2}EB}{(2\pi)^{3/2}(n m\beta)^{1/2}\Theta\sinh\big{(}\frac{\Theta B}{2E}\big{)}}  \nn \\
&& \Big{[}1-\Big{(}\frac{n\beta q E}{2m}\Big{)}^{2}\Big{]}^{-1/4} \; .\nn
\eea

Combining all the above results, the leading order QED vacuum decay rate, per unit volume at finite temperature, in the background of coexistent, homogeneous electric and magnetic fields, is given by
\be
\Gamma_{T\neq 0, \text{\tiny{fermion}}}^{E \shortparallel B}= \Gamma_{T=0, \text{\tiny{fermion}}}^{E \shortparallel B} + \Gamma_{T, \text{\tiny{fermion}}}^{E \shortparallel B} H(T-T_{*}) \; .
\ee
$\Gamma_{T=0, \text{\tiny{fermion}}}^{E \shortparallel B}$ is defined as in \Eq{eq:qedsppt0}, and $\Gamma_{T, \text{\tiny{fermion}}}^{E \shortparallel B}$ is defined as
\bwt
\bea
\Gamma_{T, \text{\tiny{fermion}}}^{E \shortparallel B} &=& \sum_{n=1}^{n_{max}} \sum_{k=0}^{\infty}2 (-1)^{k+1}\frac{q^{2}EB}{(2\pi)^{\frac{3}{2}}(nm\beta)^{\frac{1}{2}}\Theta\sinh\big{(}\frac{\Theta B}{2E}\big{)}}  \big{[}1-\big{(}\frac{n\beta q E}{2m}\big{)}^{2}\big{]}^{-\frac{1}{4}}  \cosh\Big{(}\frac{\Theta B}{2E}\Big{)}\nn \\
&& \exp\Big{[}-\frac{m^{2}}{2 q E}\big{[}2\pi (k+1) - 2\arcsin\big{(}\frac{nT_{*}}{T}\big{)}\big{]}+ \frac{nm}{2T}\sqrt{1-\frac{n^{2}T_{*}^{2}}{T^{2}}}\Big{]}\cos\Big{(}\frac{\Theta}{2}\Big{)} \; .
\eea
\ewt
Here, as before, $n_{max}=\lfloor2R/\beta\rfloor$, $H(x)$ is the Heaviside function, $T_{*}\equiv q E/2m$, and $\Theta = 2\pi (k+1)-\theta_{n}=2\pi(k+1)-2\arcsin(\frac{nT_{*}}{T})$. In the limit $B\rightarrow 0$, $\Gamma_{T\neq 0, \text{\tiny{fermion}}}^{E \shortparallel B}$ reduces to $\Gamma_{T\neq 0, \text{\tiny{fermion}}}^{E}$ and one obtains the leading order thermal corrections in QED for the pure $E$ case, thereby complementing the known result for scalars\Cit{Medina}. For $T< T_{*}$, again there are no thermal corrections and the result reverts to the $T=0$ expressions.
\section{Summary}{\label{sec:conc}}
The worldline path integral formalism provides a powerful and systematic way to compute nonperturbative vacuum decay rates in various situations. In this work, we computed leading order thermal corrections to vacuum decay rates, in SQED and QED, for the case of homogeneous, coexistent electric and magnetic fields. Apart from its theoretical importance, the results are relevant in astrophysical settings where large electric and magnetic fields may coexist in a thermal environment. 

There are a few natural avenues to follow up on that were outside the scope of the present study. The Gaussian approximation to the fluctuation prefactor is inadequate, leading to spurious singularities at thermal thresholds, and one should include higher order terms to potentially mitigate this. This is challenging even in the zero temperature case, but based on the hard thermal loop framework\Cit{Braaten:1991gm,Braaten:1992gd}, it has been argued that such spurious singularities may be softened and the result correctly interpreted\Cit{Medina}. Explicit calculation of these higher order terms beyond the Gaussian approximation would shed more light on the analytic structure of the terms at these thresholds. Another subtle point to note is that, even at zero temperature, the vacuum decay rate is not technically the same as the average, particle pair production rate\Cit{Nikishov:1970br, Cohen:2008wz}. In the zero temperature case, it may be shown that the physical observable--the mean pair production rate--is just the first term in the series for the vacuum decay rate\Cit{Nikishov:1970br}. Hence, for weak fields, the distinction is mostly pedantic. The thermal vacuum decay rates we compute are therefore expected to closely match the actual particle pair production rates for weak fields, but a more careful calculation is required to make the correspondence clear and rigorous. It would also be appealing to have a better physical understanding of the various results and reach a consensus on the remaining disagreements in the literature\Cit{Brown:2015kgj,Medina,Gould,Gould:2018ovk}. Doing away with the assumption of relatively weak fields and extending the study to arbitrary coupling strengths would also be pertinent, as well as incorporating modifications due to field inhomogeneities.


\begin{acknowledgments}
A.T would like to thank A. Brown, S. Jain  and M. Paranjape for correspondence and useful discussions. We are grateful to L. Medina and M. Ogilvie for discussions pertaining to their calculation. A.T thanks CHEP, IISc., Bangalore and DTP, TIFR, Mumbai  for hospitality, during the completion of this work.
\end{acknowledgments}

\bibliography{SPPThermalEB}

\end{document}